\documentclass[aps,prd,12pt,preprint,nofootinbib,superscriptaddress,tightenlines,floatfix,showpacs]{revtex4}
 
\usepackage{graphicx}
\usepackage{rotating}
\usepackage{dcolumn}
\usepackage{bm}
\usepackage{longtable}

\begin{document}

\newcommand{\dipi}{{\pi^+\pi^-}}
\newcommand{\dipiz}{{\pi^0\pi^0}}
\newcommand{\jpsi}{{J/\psi}}
\newcommand{\psip}{{\psi(2S)}}
\newcommand{\etap}{\eta\,'}
\newcommand{\etal}{{\sl et al.}}
\newcommand{\piz}{{\pi^0}}
\newcommand{\dilep}{{\ell^+\ell^-}}

\newcommand{\twgam}{{\gamma\gamma}}
\newcommand{\thgam}{{3\gamma}}
\newcommand{\thglu}{{3g}}
\newcommand{\gaglu}{{\gamma gg}}
\newcommand{\fogam}{{4\gamma}}
\newcommand{\figam}{{5\gamma}}
\newcommand{\gec}{{\gamma\eta_c}}
\newcommand{\cgg}{{\eta_c\to\twgam}}
\newcommand{\jthgam}{\jpsi\to \thgam}
\newcommand{\jthglu}{\jpsi\to \thglu}
\newcommand{\jgaglu}{\jpsi\to \gaglu}
\newcommand{\brjthgam}{{\cal B}(\jthgam)}
\newcommand{\brjthglu}{{\cal B}(\jthglu)}
\newcommand{\brjgaglu}{{\cal B}(\jgaglu)}
\newcommand{\jtwgam}{\jpsi\to \twgam}
\newcommand{\jgec}{\jpsi\to \gec}
\newcommand{\brjtwgam}{{\cal B}(\jtwgam)}
\newcommand{\jfogam}{\jpsi\to \fogam}
\newcommand{\brjfogam}{{\cal B}(\jfogam)}
\newcommand{\brjgec}{{\cal B}(\jgec)}
\newcommand{\brjcgg}{{\cal B}(\cgg)}
\newcommand{\bll}{{\cal B}_{\ell\ell}}
\newcommand{\jll}{\jpsi\to\dilep}
\newcommand{\brll}{{\cal B}(\jll)}
\newcommand{\bthgam}{{\cal B}_\thgam}
\newcommand{\btwgam}{{\cal B}_\twgam}
\newcommand{\bfogam}{{\cal B}_\fogam}
\newcommand{\bfigam}{{\cal B}_\figam}
\newcommand{\bthglu}{{\cal B}_\thglu}
\newcommand{\bgaglu}{{\cal B}_\gaglu}
\newcommand{\mlg}{M(\gamma\gamma)_{\rm lg}}
\newcommand{\msm}{M(\gamma\gamma)_{\rm sm}}
\newcommand{\mrec}{M(\dipi{\rm -recoil})}
\newcommand{\chip}{\chi^2/{\rm d.o.f.}}

\preprint{CLNS 08-2027}       
\preprint{CLEO 08-10}         

\title{\Large\boldmath Observation of $J/\psi\to 3\gamma$}
\author{G.~S.~Adams}
\author{M.~Anderson}
\author{J.~P.~Cummings}
\author{I.~Danko}
\author{D.~Hu}
\author{B.~Moziak}
\author{J.~Napolitano}
\affiliation{Rensselaer Polytechnic Institute, Troy, New York 12180, USA}
\author{Q.~He}
\author{J.~Insler}
\author{H.~Muramatsu}
\author{C.~S.~Park}
\author{E.~H.~Thorndike}
\author{F.~Yang}
\affiliation{University of Rochester, Rochester, New York 14627, USA}
% ...
\collaboration{CLEO Collaboration}
\noaffiliation

\date{June 3, 2008}

\begin{abstract}
We report the first observation of the decay $J/\psi\to 3\gamma$.
The signal has a statistical significance of 6$\sigma$
and corresponds to a branching fraction of 
${\cal B}(J/\psi\to 3\gamma)=(1.2\pm 0.3 \pm 0.2) \times 10^{-5}$, 
in which the errors are statistical and systematic, respectively.
The measurement uses $\psi(2S)\to\pi^+\pi^- J/\psi$ events acquired 
with the CLEO-c detector
operating at the CESR $e^+e^-$ collider.
\end{abstract}

\pacs{13.20.Gd, 12.38.Qk}

\maketitle

Ortho-positronium (oPs), the $^3S_1$ $e^+e^-$ bound state,
decays to $\thgam$ almost exclusively and has long been
a fertile ground for precision QED tests~\cite{ops}. 
The analog to oPs$\to\thgam$ for quantum chromodynamics (QCD), 
three-photon vector quarkonium decay, has not yet been observed.
The rate of three-photon $\jpsi$ decays 
acts as a probe of the strong interaction~\cite{volo},
most effectively when expressed in relation to
$\jgaglu$, $\jthglu$, or  $\jll$
due to similarities 
at the parton level.
Hence, measurements of $\bthgam$, $\bgaglu$, $\bthglu$, and $\bll$  
relative to one another
(where ${\cal B}_X\equiv {\cal B}(\jpsi\to X)$)
provide crucial experimental grounding for QCD
predictions~\cite{volo,qcd,kwong}.

  In this Letter we report the first observation of $\jthgam$.
Rate measurements for other rare or forbidden all-photon decays, 
$\jtwgam$, $\fogam$, $\figam$, and $\gec$ with $\cgg$,
are also described.
Previous searches for $\omega$ and $\jpsi$ decay to $\thgam$ have yielded 
branching fraction upper limits 
of 1.9$\times$10$^{-4}$ and 5.5$\times$10$^{-5}$, respectively~\cite{PDG2006}. 
As with oPs, {$C$-parity} symmetry suppresses vector quarkonia decays to
an even number of photons, and two-photon
decays are forbidden by Yang's theorem~\cite{yang}.
Ref.~\cite{bes:2gamma} reports the limit 
$\btwgam$$<$2.2$\times$10$^{-5}$ at 90\% confidence level (C.L).
Five-photon decays are suppressed by an
additional factor of (at least) $\sim\alpha^2$; {\sl c.f.,}
${\cal B}({\rm oPs}\to 5\gamma)\sim 2\times 10^{-6}$~\cite{orthopfivegamma}.

Ignoring QCD corrections altogether,
Ref.~\cite{kwong} predicts  
$\bthgam/\bll\approx \alpha/14$,
$\bthgam/\bgaglu\approx (\alpha/\alpha_s)^2/3$ and 
$\bthgam/\bthglu \approx (\alpha/\alpha_s)^3$. 
Using the precisely measured $\bll$~\cite{PDG2006} 
in the first prediction implies $\bthgam\approx 3\times$10$^{-5}$.
The latter two suffer the uncertainty of what 
value of $\alpha_s$ to employ at the charmed quark mass scale~\cite{volo}.
Assuming $\alpha_s(m_c^2)$=0.3 and inserting the result
from a recent CLEO measurement~\cite{besson} 
($\bgaglu\approx 0.09$ and $\bthglu\approx 0.66$) into the
latter two predictions gives $\bthgam \approx (0.9$\,-$1.6)\times$10$^{-5}$.
The first-order perturbative QCD corrections~\cite{kwong} to these
estimates are large, so these predictions 
should only be considered as approximate. 

 Events were acquired at the CESR 
$e^+e^-$ collider with the CLEO detector~\cite{CLEO}, 
mostly in the CLEO-c configuration (95\%) with the balance
from CLEO~III. The dataset
corresponds to 27$\times$10$^6$ produced 
$\psip$ mesons and (9.59$\pm$0.07)$\times$10$^6$ 
$\psip\to\dipi\jpsi$ decays~\cite{xnext}. 
Event selection requires the tracking system to find exactly two 
oppositely charged particles, corresponding to the $\dipi$
recoiling from the $\jpsi$, and that the
calorimeter have at least 2, 3, 4, 5, and 3
photon showers for the $\jtwgam$, $\thgam$, $\fogam$, $\figam$,
and $\gec(\to\twgam)$ samples, respectively.
Photon candidates must have energy exceeding 36~MeV 
and, with respect to any shower associated with one 
of the charged pions, either be located $(a)$ more
than 30~cm away, or $(b)$ between 15~cm and 30~cm from it
{\it and} have a photon-like lateral shower profile. 
We require that photon candidates not be located 
near the projection of either pion's trajectory 
into the calorimeter nor be aligned with the initial momentum 
of either pion within 100~mrad.

A two-step kinematic fit first constrains the beam spot 
and the two charged pion
candidates to a common vertex, and then the
vertexed $\dipi$ and the most energetic $n$ photon candidates to the 
$\psip$ mass~\cite{PDG2006} and initial 
three-momentum, including the effect of the 
$\simeq$3~mrad crossing angle between the
$e^+$ and $e^-$ beams. Tight quality restrictions are
applied to the vertex ($\chi^2_{\rm v}$/d.o.f.$<$3) and 
four-momentum ($\chip$$<$3) fits. 
The mass recoiling against the $\dipi$ must lie inside
a window around the $\jpsi$ mass,
$\mrec =$3087-3107~MeV. Non-$\jpsi$
backgrounds are estimated by keeping 
a separate tally of events with $\mrec$
inside 2980-3080~MeV or 3114-3214~MeV, ranges which
together are ten times wider than the signal window. 

Events with {\sl any} of the photon pairs in the mass windows 
0.10-0.16~GeV, 0.50-0.60~GeV, or 0.90-1.00~GeV
are rejected to eliminate contributions from decays with
$\piz$'s, $\eta$'s, or $\etap$'s, the dominant
sources of photons in $\jpsi$ decays.
For the 3$\gamma$ selection only, we require
all photon pair masses be less than 2.8~GeV to
eliminate potential contamination from $\eta_c\to\gamma\gamma$. 
This requirement effectively restricts 
the smallest energy photon to have energy exceeding 
200~MeV. For the 4$\gamma$ and 5$\gamma$ samples only,
the smallest shower energy must be above 120~MeV,
and all lateral shower profiles must be photon-like.
This last restriction on shower shape avoids 
feed-up from $\jpsi\to\gamma\eta(')$, $\eta(')$$\to$$\gamma\gamma$ 
events with one or more photon conversions
between the tracking chambers and the calorimeter:
in such cases the two showers from the conversion $e^+$
and $e^-$ overlap one another, thereby distorting
both of their lateral profiles.
For the $\gamma\eta_c$ channel only,
we restrict the search region to large 
$\mlg$ and small $\msm$, which 
are, respectively, the largest and smallest
of the three two-photon mass combinations
in the event. The signal region is chosen this way so as
to keep backgrounds small. Specifically, the signal box
is defined, in units of GeV, 
by 0.16$<$$\msm$$<$0.48, 2.985$<$$\mlg$$+$$0.0935\msm$$<$3.040.

Signal and background decay modes are modeled 
with Monte Carlo (MC) samples that were generated using the
{\sc EvtGen} event generator~\cite{evtgen},
fed through a {\sc Geant}-based~\cite{geant} detector simulation,
and then exposed to event selection criteria. 
For $\jpsi\to n\gamma$ signal decays, final state photon momenta are distributed
according to phase-space. For $\jthgam$, the
lowest order matrix element for ortho-positronium~\cite{ore} is used
as an alternate; compared to phase space, it modestly
magnifies the configurations that are
two-body-like and those with 
three nearly-equal-energy photons (at the expense
of topologies lying between these two extremes).
For the process $\jgec$, an $\eta_c$ mass and width of
2979.8~MeV and 27~MeV, respectively, are used (both
are close to the PDG values~\cite{PDG2006}) to generate
a Breit-Wigner $\twgam$-mass distribution; alternate
widths from 23-36~MeV and different lineshapes~\cite{ryan} are
explored as systematic variations.

\begin{figure}[tbh]
\begin{center}
\includegraphics[width=5.5in]{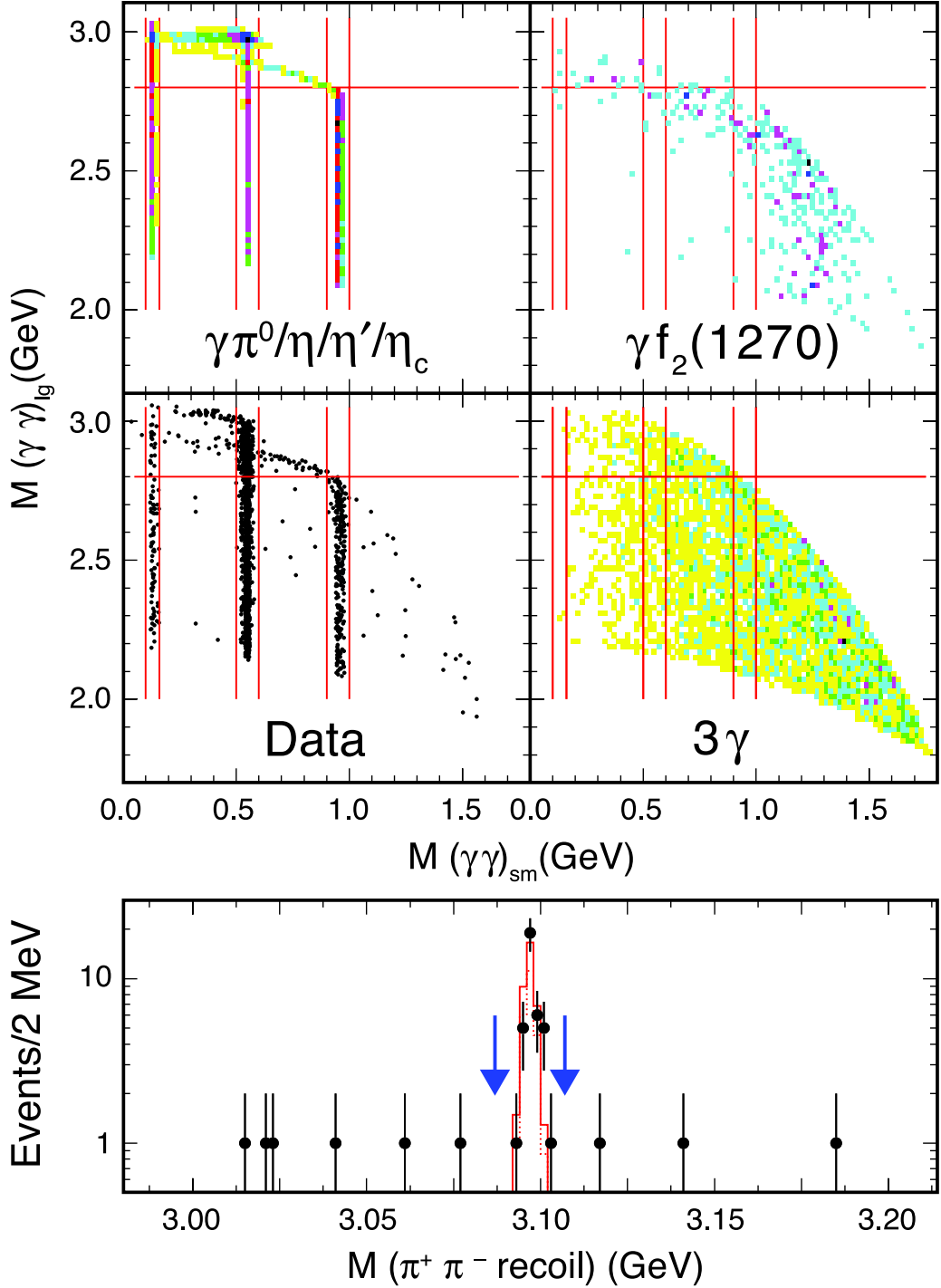}
\caption{Top four plots:
in $3\gamma$ data (lower left) and MC events for
different $\jpsi$ decays
(top row and lower right), the largest {\sl vs.~}the
smallest two photon mass combination per event.
In the MC plots, darker shading of each bin
signifies higher event density than lighter shading;
in the data plot, each dot represents an event.
The solid lines demarcate regions excluded from
the $\jthgam$ selection.
Bottom plot: distribution of $\mrec$
for the data events (points with error bars)
overlaid with the $\jthgam$ signal MC prediction (dotted line histogram)
and MC background plus signal (solid line histogram)
normalized to the data population. The arrows indicate
the region of accepted recoil mass.
}
\label{fig:fig1}
\end{center}
\end{figure}

\begin{figure}[tbh]
\begin{center}
\includegraphics[width=5.5in]{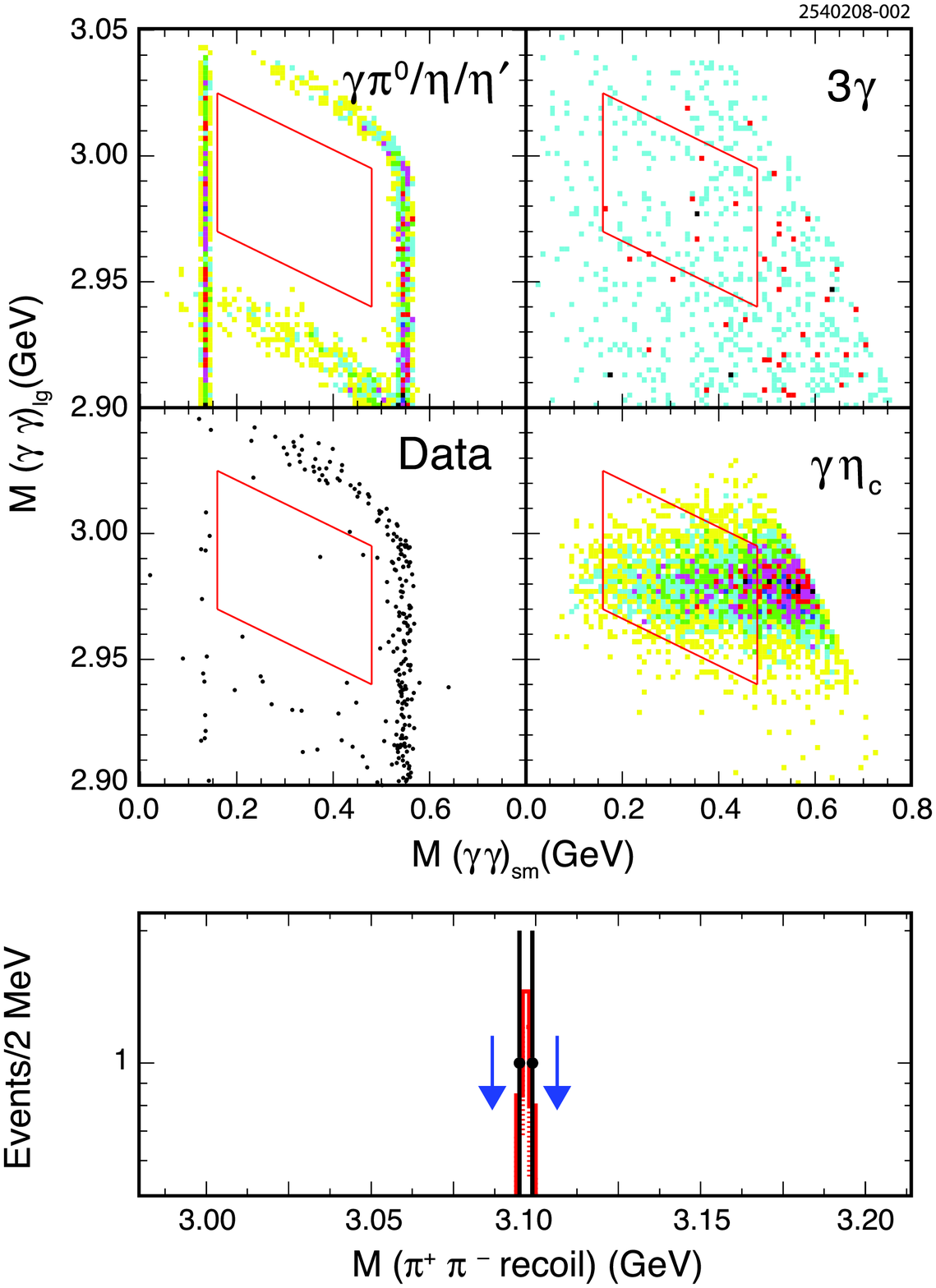}
\caption{As in Fig.~\ref{fig:fig1}, except 
zoomed in on the $\eta_c$ region, and
the overlaid parallelogram indicates the
$\jgec$ signal region.
 The 
$\mrec$ distribution is for
$\gec$ candidates.
}
\label{fig:fig2}
\end{center}
\end{figure}

Distributions in $\msm$ vs.~$\mlg$ and $\mrec$ for the $\jthgam$
and $\jgec(\twgam)$ samples are shown for data, signal MC samples,
and likely background decays in Figs.~\ref{fig:fig1} and \ref{fig:fig2},
respectively.

\begin{figure}[thb]
\begin{center}
\includegraphics[width=6.0in]{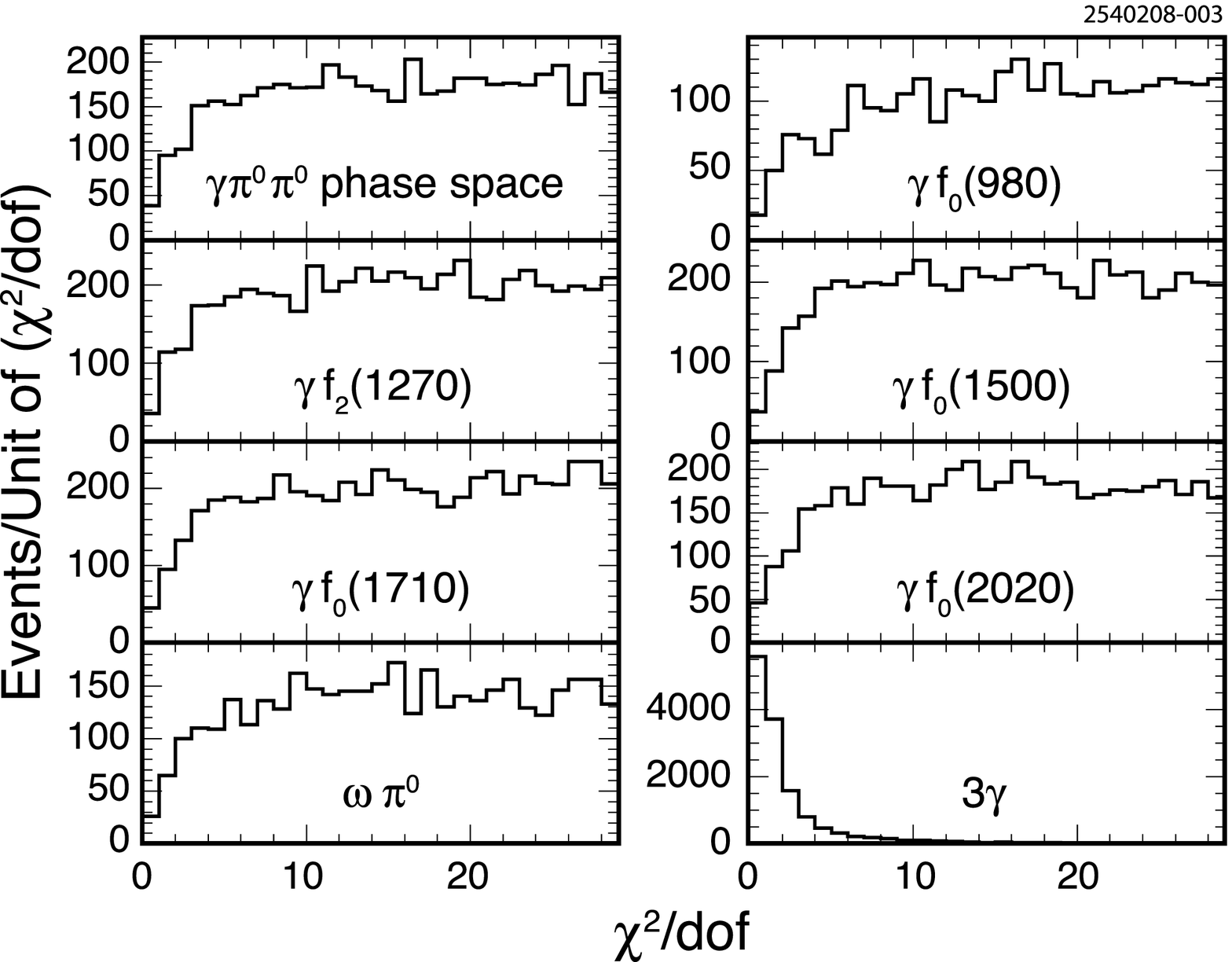}
\caption{The distribution of $\chip$
for $\jthgam$ (lower right) and several sources
of $\gamma\piz\piz$ background.
}
\label{fig:fig3}
\end{center}
\end{figure}

\begin{figure}[t]
\begin{center}
\includegraphics[width=5.2in]{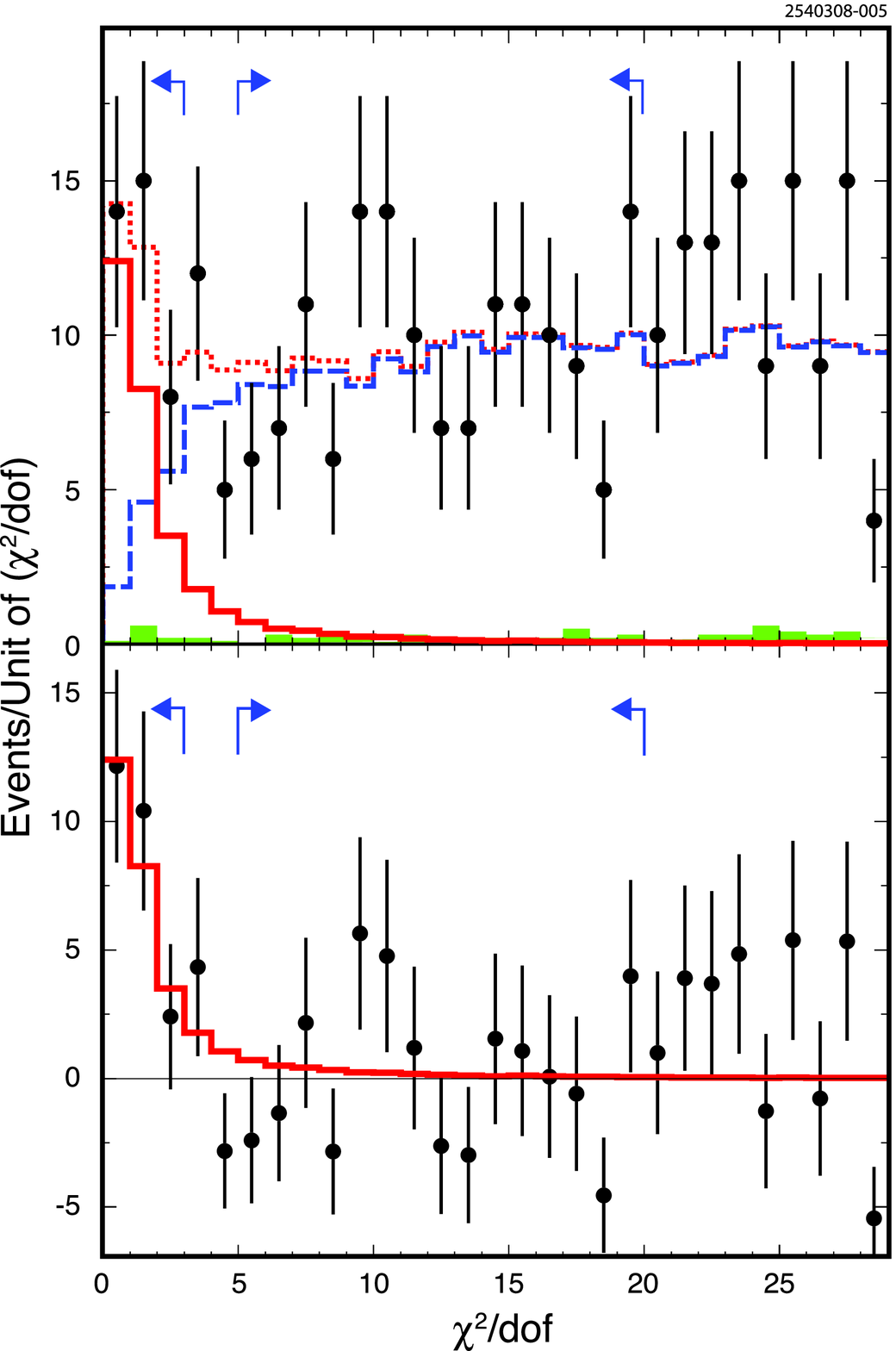}
\caption{The distribution of $\chip$
for $\jthgam$, in the top plot showing data (points with error bars) 
overlaid with the sum (dotted line histogram) of three components:
non-$\jpsi$ background from scaled data sidebands (shaded histogram)
and MC predictions for signal (solid) and $\jpsi\to\gamma\piz\piz$ 
background (dashed). The bottom plot shows
the same distribution, but with the MC and non-$\jpsi$ background
subtracted from the data. The arrows indicate the values
for signal selection and background normalization.
}
\label{fig:fig4}
\end{center}
\end{figure}

In all modes, non-$\jpsi$ backgrounds are small and are 
subtracted statistically
using $\mrec$ sidebands in the data.
We determine the backgrounds from $\jpsi$ decays
with an exhaustive study of Monte Carlo samples.
Decays with $\jpsi\to \gamma f_J$
(where $f_J$ signifies any of the many
isoscalar mesons in the mass range from 600-2500~MeV),
followed by $f_J\to\twgam$ pose a negligible threat 
for any of the target modes because
the product branching fractions are extremely small
({\sl e.g.}, $\simeq$2$\times$10$^{-8}$ for $\jpsi\to\gamma f_2(1270)$,
$f_2(1270)\to\gamma\gamma$).
The predominant source of backgrounds to the 3$\gamma$
sample is the $\gamma\piz\piz$
final state. This type of event can survive the selection by
having both $\piz$ decay axes nearly parallel to 
their lines of flight, such that one photon of each pair has very low
energy in the laboratory frame, 
and is therefore
nearly irrelevant to conservation of four-momentum.
An analysis by BES~\cite{bes:5gamma} found that the largest sources
of $\jpsi\to\gamma\piz\piz$ are from $\jpsi\to\gamma f_J$ decays,
specifically through $f_2(1270)$ and $f_0(2050)$, followed in
importance by $f_0(1710)$, $f_0(1500)$, and a
number of much smaller contributions from nearby resonances.
However, not all relevant product branching fractions 
for $\jpsi\to\gamma f_J$, $f_J\to\piz\piz$ have been 
measured, those that are measured have 
large uncertainties, and interference effects
among overlapping $f_J$ may not be small. 
A method to normalize $\gamma\piz\piz$ 
{\sl other} than using measured branching fractions 
is employed to reduce systematic uncertainty. The $\chip$
distribution for $\gamma\piz\piz$ decays has a characteristic
shape, {\sl nearly independent of $\piz\piz$ mass}, 
as shown in Fig.~\ref{fig:fig3}: the region
$\chip$=5-20, where almost no signal is present, 
is used to establish the level of $\jpsi\to\gamma\piz\piz$.
Figure~\ref{fig:fig4} shows the $\chip$ distribution 
from data, MC signal and MC background and the
small contribution from non-$\jpsi$ decays obtained
from the $\mrec$ sidebands.
The 142 data events with $\chip$=5-20 contain $\jthgam$ signal (3.4 events),
non-$\jpsi$ background (3.2), and, using known branching fractions,
$\jpsi\to\omega\eta$, $\eta\to\twgam$ (1.7),
$\jpsi\to\gamma\eta$, $\eta\to\twgam$ (1.2),
$\jpsi\to\gamma\eta$, $\eta\to 3\piz$ (1.2),
$\jpsi\to\gamma\etap$, $\etap\to \gamma\omega$, $\omega\to\piz\gamma$ (0.6),
and $\jpsi\to\gamma\piz$ (0.2). The remainder (130.5 events)
serves to normalize the $\gamma\piz\piz$ background component,
which has a relative 8\% statistical uncertainty. 

As a cross check on the $\thgam$ background normalization,
we perform a maximum likelihood fit to data in
the entire $\jthgam$ $\chip$=0-30 region with 
the combination of shapes from MC of $\gamma\piz\piz$
and $\thgam$ signal with floating normalizations for each,
and a fixed $\jpsi$-sidebands contribution from data,
scaled by a factor of 0.1. Using this method with 
different sources of the $\gamma\piz\piz$
taken one at a time as 100\% of the background
results in an average signal size of 23.3 events
(with variation from 22.8 to 24.1), which is 0.9
events smaller than our nominal technique.
Based on these numbers we assign 
a systematic error of 0.9 events, or $\simeq$5\% relative, 
for signal extraction and background estimation for $\jthgam$.

The $\chip$ fit just described is repeated with 
the $\thgam$ signal shape weight fixed to zero. The 
likelihood difference with respect to the nominal
fit provides a measure of the statistical
significance of the signal. This significance varies
from 5.9$\sigma$ to 6.6$\sigma$ when using any one of the
backgrounds $\gamma f_2(1270)$, $\gamma f_0(1500)$,
 $\gamma f_0(1710)$, $\gamma f_0(2020)$, $\gamma \piz\piz$ (phase space) 
as the sole contributor to the background shape.

MC studies indicate the following primary sources of backgrounds
for the other modes: for the 2$\gamma$ sample,
$\jpsi\to\gamma\piz$ (3.3 events) and $\gamma\eta$, $\eta\to\twgam$ (2.7);
for the 4$\gamma$ sample,
$\jpsi\to\gamma\eta$, $\eta\to\twgam$ (0.9) or 
$\eta\to 3\piz$ (0.8),
$\gamma\etap$, $\etap\to\twgam$ (0.3) or 
$\etap\to\gamma\omega$, $\omega\to\piz\gamma$ (0.9) or
$\etap\to\piz\piz\eta$, $\eta\to\twgam$ (0.3); 
for the 5$\gamma$ sample,
$\jpsi\to\gamma\eta$, $\eta\to 3\piz$ (0.2)
and $\gamma\etap$, $\etap\to\piz\piz\eta$, $\eta\to\twgam$ (0.3);
for $\gamma\eta_c$, 
$\jpsi\to\gamma\eta$, $\eta\to\twgam$ (0.3), 
$\gamma\etap$, $\etap\to\twgam$ (0.2),
and our newly found signal, $\jpsi\to 3\gamma$ (0.3).

Numerical results appear in Table~\ref{tab:results}.
Net yield uncertainties and upper limits on event counts include
the effects of statistical fluctuations in signal and background estimates.
Signal efficiencies range from $\simeq$2\% ($\figam$) to
$\simeq$22\%\ ($\thgam$), and $\jthgam$ is the only mode
with a clear signal: 37 events observed on a background of
12.8.
Statistics dominate the overall uncertainties for all decay modes.
The $\jthgam$ efficiencies for pure phase-space
and the oPs matrix element are equal to within (0.2$\pm$0.1)\%;
nevertheless, a 15\% systematic error is assigned
to allow for different behavior in the much heavier
$\jpsi$ system. For $\gamma\eta_c$, uncertainties in the lineshape,
background, and $\Gamma(\eta_c)$ dominate the systematic error.

Using the recently determined 
${\cal B}(\jpsi\to\gamma\eta_c)$= (1.98$\pm$0.09$\pm$0.30)\%~\cite{ryan}, 
the $\eta_c\to\twgam$ branching fraction can be calculated as
${\cal B}(\eta_c \to \twgam)$=(0.6$^{+1.3}_{-0.5}\pm$0.1)$\times$10$^{-4}$, or 
$<$3$\times$10$^{-4}$ at 90\%~C.L.
This value is consistent with the PDG~\cite{PDG2006} 
fit value of (2.7$\pm$0.9)$\times$10$^{-4}$
at the level of 1.3$\sigma$, although making
a meaninful comparison is difficult because the PDG
number depends indirectly upon previous, considerably smaller
values for ${\cal B}(\jpsi \to\gamma\eta_c)$.

In conclusion, we have investigated decays $\jpsi \to n\gamma$ 
with $n=2,3,4,5$, where the photons are produced in direct decay, 
not through an intermediate resonance. For $n$=3, a signal of 
6$\sigma$ significance is found with branching fraction 
$\bthgam$=(1.2$\pm$0.3$\pm$0.2)$\times$10$^{-5}$.
This value lies between the zeroth order
predictions~\cite{kwong} for $\bthgam/\bgaglu$ 
and $\bthgam/\bthglu$ and is consistent 
with both,
but is a factor of $\simeq$2.5 below that of $\bthgam/\bll$.
This measurement represents the first observation of 
a three-photon meson decay.
No signal is seen for $n$=2, 4, or 5, and upper limits are set
on the branching fractions,
each of which is the most precise or only measurement.
We also measure 
${\cal B}(\jpsi \to \gamma\eta_c)\times {\cal B}(\eta_c\to\twgam)$=
(1.2$^{+2.7}_{-1.1} \pm$0.3)$\times$10$^{-6}$ or
an upper limit of $<$6$\times$10$^{-6}$ at 90\%~C.L., 
both consistent with other determinations~\cite{PDG2006}.

We gratefully acknowledge the effort of the CESR staff
in providing us with excellent luminosity and running conditions.
This work was supported by
the A.P.~Sloan Foundation,
the National Science Foundation,
the U.S. Department of Energy,
the Natural Sciences and Engineering Research Council of Canada, and
the U.K. Science and Technology Facilities Council.

\begin{table*}[tb]
\setlength{\tabcolsep}{0.50pc}
\catcode`?=\active \def?{\kern\digitwidth}
\caption{ Results for the five $\jpsi\to n\gamma$ decay modes,
showing the raw number of
signal candidate events, estimated background levels, 
statistical significance of each signal, the net event yield and its
90\%~C.L. upper limit (UL), the signal efficiency,
different sources of systematic error and their
quadrature sum, expressed
in percent of the central value ($\thgam$, $\gec$)
or of the UL (others), the branching
fraction 
${\cal B}(\jpsi\to X)$ [first error is statistical, second is systematic],
and the corresponding 90\%~C.L.~upper limit UL,
including effects of systematic errors.
}
\label{tab:results}
\begin{center}
\begin{tabular}{lccccc}
\hline
\hline
          & $2\gamma$ & $3\gamma$ & $4\gamma$ & $5\gamma$ &
          $\gamma\eta_c, \eta_c\to\gamma\gamma$\\
\hline
Signal candidates (events)      &  9   &  37  & 5   & 0   & 2   \\ 
Background estimates (events) & & & & & \\
\qquad $\jpsi$ backgrounds      & 6.2  & 11.9 & 3.2 & 0.5 & 0.8 \\
\qquad Non-$\jpsi$ backgrounds  & 0.9  & 0.9  & 0.5 & 0   & 0   \\
Background sum (events)         & 7.1  & 12.8 & 3.7 & 0.5 & 0.8 \\ 
& & & & & \\
Statistical significance ($\sigma$) & 1.1 & 6.3 & 1.0 & 0.0 & 1.0 \\
Net yield (68\%~C.L. interval) (events) & 1.9$^{+4.7}_{-1.6}$  &  24.2$^{+7.2}_{-6.0}$ & 1.3$^{+2.4}_{-1.3}$    & 0$^{+1.2}_{-0}$ &$1.2^{+2.8}_{-1.1}$\\
UL @~90\%~C.L. (events)        & $<$7.7 & $<$33.5               & $<$6.0 & $<$2.3 &$<$4.7\\
Efficiency (\%)    & 19.2 & 21.8& 8.71& 1.90& 10.9\\
Systematic uncertainties (\%) & & & & & \\
\qquad Matrix element          &   0  & 15  & 15  & 15  & 15  \\
\qquad $\jpsi$ background      &  15  & 5   & 10  &  0  & 15  \\
\qquad $\dipi\jpsi$ counting   &  0.7 & 0.7 & 0.7 & 0.7 & 0.7 \\
\qquad Detector modeling       &  4.5 & 6.4 & 8.3 & 10  & 6.4 \\
\qquad $\Gamma(\eta_c)$        &    0 &  0  &  0  &  0  & 12  \\  
Quadrature sum (\%)            &  16  & 17  & 20  & 18  & 25  \\ 
& & & & & \\
${\cal B}(\jpsi\to X)$ [$10^{-6}$] & & $12\pm 3\pm 2$&& & 1.2$^{+2.7}_{-1.1}\pm 0.3$ \\
UL on ${\cal B}(\jpsi\to X)$ @~90\%~C.L. [$10^{-6}$] & $<$5 & $<$19& $<$9 & $<$15  & $<$6 \\
\hline
\hline
\end{tabular}
\end{center}
\end{table*}
 
\end{document}